\documentclass[preprint,3p]{elsarticle}

\usepackage{amssymb}
\usepackage{soul}

\usepackage{caption}
\usepackage{url}
\usepackage{natbib}

\setcitestyle{authoryear,open={(},close={)}}

\journal{peer review.}

\begin{document}

\begin{frontmatter}



\title{Writing about COVID-19 vaccines: Emotional profiling unravels how mainstream and alternative press framed AstraZeneca, Pfizer and vaccination campaigns}





\author[a]{Alfonso Semeraro}
\author[a]{Salvatore Vilella} 
\author[a]{Giancarlo Ruffo} 
\author[b]{Massimo Stella}

\affiliation[a]{organization={Department of Computer Science, University of Turin},
            city={Turin},
            country={Italy}}

\affiliation[b]{organization={CogNosco Lab, Department of Computer Science, University of Exeter},
            city={Exeter},
            country={UK}}

\begin{abstract}

Ever since their announcement in November 2020, COVID-19 vaccines were largely debated by the press and social media. With most studies focusing on COVID-19 disinformation in social media, little attention has been paid to how mainstream news outlets framed COVID-19 narratives compared to alternative sources. To fill this gap, we use cognitive network science and natural language processing to reconstruct time-evolving semantic and emotional frames of 5745 news about COVID-19 vaccines. Our dataset covers 17 outlets over 8 months and includes Italian news articles that were massively re-shared on Facebook (5 mil. total shares) and Twitter (200k total shares). We found consistently high levels of trust/anticipation and less disgust in the way mainstream sources framed the general idea of "vaccine/vaccino". These emotions were crucially missing in the ways alternative sources framed COVID-19 vaccines. More differences were found within specific instances of vaccines. Alternative news included titles framing the AstraZeneca vaccine with strong levels of sadness, absent in mainstream titles. Mainstream news initially framed "Pfizer" along more negative associations with side effects (e.g. "allergy", "reaction", "fever") than "AstraZeneca". With the temporary suspension of the latter vaccine, on March 15th 2021, we identified a semantic/emotional shift: Even mainstream article titles framed "AstraZeneca" as semantically richer in negative associations with side effects, while "Pfizer" underwent a positive shift in valence, mostly related to its higher efficacy. \textit{Thrombosis} entered the frame of vaccines together with fearful conceptual associations, while the word \textit{death} underwent an emotional shift, steering towards fear in alternative titles and losing its hopeful connotation in mainstream titles, with the lack of anticipation. Our findings expose crucial aspects of the emotional narratives around COVID-19 vaccines adopted by the press, highlighting the need to understand how alternative and mainstream media report vaccination news.


\end{abstract}



\begin{keyword}
natural language processing \sep text analysis \sep complex networks \sep cognitive network science \sep COVID-19 \sep COVID-19 vaccines
\end{keyword}

\end{frontmatter}


\section{Introduction}

Vaccination campaigns are quickly turning the table against COVID-19. Massive efforts were put in place by most countries to acquire and distribute millions of doses all over the world~\citep{rolland2021covid19}. Ever since their announcement in November 2020, vaccines were largely covered, described and debated by news and social media, creating a deluge of information consumed by individuals~\citep{puri2020social,stella2021vaccines,murphy2021psychological}. 

Whereas many studies focused on the structural and dynamical features of COVID-19 knowledge flows in social media~\citep{castioni2021voice,puri2020social,jiang2021social}, less well explored is the other half of news media consumption, represented by constellations of highly credible/mainstream and lowly credible/alternative journal venues~\citep{bridgman2020causes,cinelli2020covid}. Newspapers not only convey succinct information about the happening of events but can often bolster awareness about specific aspects of events~\citep{apuke2021fake,gozzi2020collective}, e.g., bolstering the fatal consequences of statistically rare side effects of vaccines, or promote specific emotional perceptions~\citep{stella2021cognitive}, e.g., painting the announcement of vaccine shortages with concern or hopefulness for the future. 

Ultimately both social media and newspapers represent key components of information consumption~\citep{castioni2021voice,vilella2021impact}: they promote knowledge and specific perceptions about real-world events like vaccines. Exploring the semantic and emotional profiles of knowledge disseminated by such venues becomes therefore essential to reconstruct key ideas read by and influencing massive audiences~\citep{stella2021cognitive}. Specifically for COVID-19 vaccines, this reconstruction is urgently needed to rethink how different news outlets, over time, structured knowledge around vaccinations that reached massive audiences~\citep{bridgman2020causes}. While monitoring the presence of emotions like anger or trust in massively re-shared knowledge is important~\citep{yang2021covid}, the main challenge to understand how these emotions can affect individuals is finding out which are the main concepts eliciting such emotions in massively read content~\citep{sulis2016figurative,stella2021cognitive}. Identifying the specific emotional and conceptual associations promoted by news media remains a crucial achievement for fighting misinformation and social manipulation~\citep{cinelli2020covid}.

This manuscript adopts an interpretable natural language processing (NLP) framework of narratives centered around COVID-19 vaccines, promoted by mainstream/alternative media outlets subsequently re-shared on social media like Twitter and Facebook. We select the Italian news system as a complex yet relatively unexplored case study (cf.~\citealt{stella2021vaccines}).  The current investigation adopts the recent frameworks of cognitive network science~\citep{siew2019cognitive} and forma mentis networks~\citep{stella2021cognitive} to interpret language processing and unveil the structure of knowledge embedded in news articles as syntactic/semantic networks of conceptual associations. Adopting semantic frame theory from psycholinguistics~\citep{fillmore2006frame}, we reconstruct the meaning attributed to vaccines and other terms in language by looking at their semantic and emotional associations. Finding these associations operationalises a model for meaning reconstruction from text, a cognitive task where networks of conceptual associations between concepts cast meaning and emotional context to each individual concept~\citep{fillmore2006frame,carley1993coding}. This approach has been commonly used as content mapping through human coding~\citep{carley1993coding}, which is clearly impractical to analyse thousands of news papers. The methodology outlined in this manuscript describes how artificial intelligence methods can automatise meaning reconstruction through cognitive networks and thus parse large volumes of texts at once.

Reconstructing the semantic and emotional frames surrounding COVID-19 vaccines is fundamental for understanding which perceptions pressured individuals taking part in vaccination campaigns. A lack of trust towards something can crucially inhibit adherence to norms promoted by institutions, slowing down vaccination enrolment. Similarly, semantic associations linking ``vaccine'' with ``hoax'' or ``conspiration'' could bolster conspiracy theories, altering the risk-perception of individuals and ultimately exposing them to contagion. Despite the analysis of these patterns often faces the complexity of trends and behaviours present in social media data, that might be unaffected by flickering emotions or contrasting content~\citep{valensise2021lack}, the scientific community must consider richer semantic and emotional maps of knowledge exchange to better understand how social media impact real-world behaviours. 
  
The manuscript is organised in the following way. We first review relevant works that motivate our cognitive network approach to reconstructing vaccine perceptions in alternative and mainstream news media. Our results interestingly unveil drastic differences in the ways alternative and mainstream outlets framed COVID-19 vaccines within their news titles. We also detect structural shifts in the semantic frames of AstraZeneca and Pfizer, two specific types of COVID-19 vaccines. Our cognitive networks also highlighted the emergence of strong debates in news about the side effects of specific types of vaccines: Not all COVID-19 vaccines received the same treatment/semantic framing from alternative and mainstream news. We discuss our results in light of related literature and outline our machine learning methods at the end of the manuscript. 

\begin{figure*}[ht!]
    \centering
    \includegraphics[width=\textwidth]{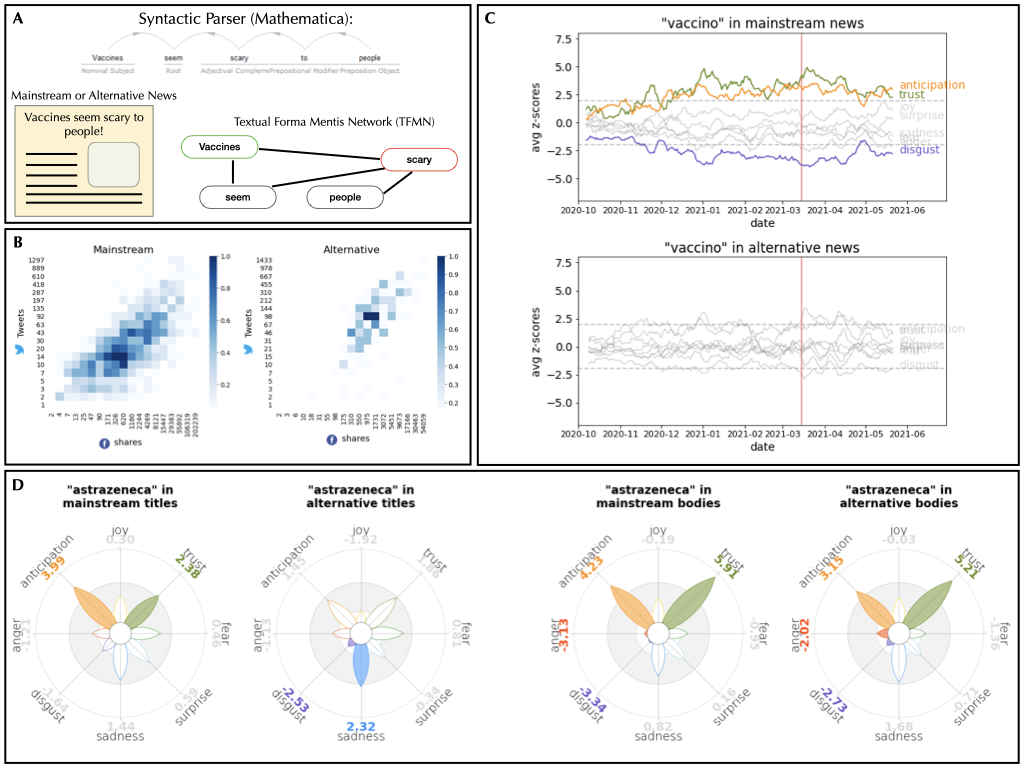}
    
      \caption{(\textbf{A}) Mechanisms of data creation and key differences highlighted by them in mainstream and alternative news. Data is grouped by weeks. Figure 1 (a) highlights how our approach gives structure to knowledge in news. --- (\textbf{B}) Each tile $<i, j>$ is coloured after the number of urls that have been shared $i$ times on Facebook and $j$ times on Twitter. Both axes are logarithmically binned.--- (\textbf{C}) Z-scores of emotions in news that contain the word ``vaccin'' against a neutral sample, grouped by day and smoothed with a weekly rolling average. Mainstream news about the word ``vaccin'' show consistent high levels of Trust and Anticipation, conveying hope for the vaccination campaign, and significantly less Disgust. This positive leaning is not visible in alternative news. ---(\textbf{D})  Distribution of z-scores of bodies (right) and titles (left) into the emotional frame of the word ``astrazenec'', divided by mainstream and alternative outlets. There is a striking difference in how titles are written by mainstream and alternative news outlets, with the latter evoking more sadness but also significantly less disgust than the former. Such a difference is not visible in the articles bodies. White filled petals represent emotions that are not significantly over- or under-represented in the corpus, if compared with a neutral baseline.}
    \label{fig:vaccin_daily}
\end{figure*}

\subsection{Related Works}

Humans can communicate their ideas through language~\citep{aitchison2012words}, which makes it key to use linguistic data for mapping perceptions and attitudes. Both in computer science and psycholinguistics, the problem of detecting positive or negative perceptions from language is known as \textit{stance detection}~\citep{stancedetectionsurvey}. Although historically psycholinguistic focused on human coding of texts~\citep{biber1989styles}, in the last decade several approaches from artificial intelligence in computer science have brought to accurate predictions of stances in texts via machine learning techniques like deep neural networks~\citep{kuccuk2020stance,zarrella2016mitre} or sentiment analysis~\citep{rudkowsky2018more}. A key limitation of these approaches is their black-box nature, where the experimenter cannot easily interpret the inner structure of the model fitted from the trained data and its relationship with the specific stance under scrutiny~\citep{rudin2019stop}. To overcome this limitation, a novel stream of stance detection approaches adopts complex networks as tools for representing and visualising key aspects of text-based discourse~\citep{saif2012semantic,radicioni2021analysing,stella2018bots,lai2019stance,lai2020stance}. Complex networks have the distinctive advantage of highlighting topological patterns of connectivity between interconnected entities, be it online users engaging in replies/mentions/re-sharing~\citep{vilella2020immigration,radicioni2021analysing} or specific hashtags or words co-occurring together within the same text~\citep{akimushkin2017text,amancio2015probing,stella2018bots}. Complex networks represent a powerful tool for enriching machine learning approach and highlighting the inner structure of a given stance promoted by online users, i.e., for understanding how ideas were associated and framed by online audiences~\citep{stella2021cognitive}.

Complex networks of conceptual associations can also give stances a measurable structure, highlighting how ideas/concepts/emotions were associated and framed by specific textual narratives~\citep{stella2021forma,de2019paragraph}. Among many options, associations between concepts can be reconstructed mainly in three ways: through word co-occurrences~\citep{de2019paragraph,amancio2012structure}, e.g., detecting adjacent words in sentences; (ii) through semantic associations~\citep{kenett2017semantic,akimushkin2017text}, e.g., words being synonyms or reminding of each other; or (iii) through syntactic relationships, e.g., words specifying the meaning of each other. Since word co-occurrences can approximate syntactic relationships once words devoid of meaning are filtered out~\citep{cancho2001small,amancio2015probing}, the above methods can transform unstructured texts into networks of interconnected words/concepts. These structures can then be enriched with emotional data, identifying the key emotions~\citep{mohammad2013,mokryn2020sharing} or sentiment patterns~\citep{mohammad2016sentiment} evoked by concepts. The resulting network structure is indicative of associative knowledge and emotional signatures embedded in texts~\citep{stella2020forma}, partially reflecting the way stances were organised in the authors' psychology~\citep{jackson2020text}. In fact, both the frameworks of content mapping in communication science~\citep{carley1993coding} and frame semantics in cognitive psychology~\citep{fillmore2008frame} indicate that the way words are linked in semantic/syntactic networks provides key insights for reconstructing the meaning attributed to words in texts by authors.

This relationship between network structure and meaning in narratives has been used for a variety of automatic tasks. Amancio and colleagues used co-occurrence networks of words to perform author identification of novels~\citep{amancio2015probing,amancio2012structure}. Stella and colleagues~\citep{stella2018bots} used hashtag co-occurrence networks to highlight the negative, hatred-inspiring stances injected by automated accounts on social discourse about the Catalan referendum in 2018. Colladon~\citep{colladon2018semantic} combined multiple network metrics of word-word co-occurrences in social discourse to identify key emotional and semantic features of news relative to brand advertisement. Ferrara and colleagues~\citep{ferrara2015quantifying} found that in social media, textual posts richer in positive emotions could reach larger audiences whereas a faster spreading rate was found for stances richer in negative emotional content. Radicioni and colleagues~\citep{radicioni2021analysing} combined hashtag co-occurrences and online social interactions to identify discursive communities engaging in different stances of immigration. The authors identified a strong polarisation in favour and against humanitarian interventions to immigration mainly correlated with political factions, in agreement with previous social network approaches~\citep{vilella2020immigration}. Teixeira and colleagues~\citep{teixeira2021revealing} used semantic/syntactic networks to reconstruct stances expressed in suicide notes and found negative perceptions of concepts like ``love'', distorted by suicide ideation when compared to control data. Mokryn and colleagues found that the emotional words expressed in movie reviews were predictive of the emotions inspired by those movies, further strengthening a connection between textual data and cognitive/emotional content.

Although differing in their scope and methods, the above approaches have a key common element: They all perform quantitative measurements of the semantic, syntactic and emotional content of texts. This automation makes it possible to perform stance detection through volumes of data/texts that would be intractable with human coding~\citep{biber1989styles}.

In this work, we build upon the above past approaches by reconciling interpretable machine learning and networks of conceptual associations within the framework of textual forma mentis networks (TFMN)~\citep{stella2020forma}. These networks perform dependency parsing - powered by recurrent neural networks~\citep{dozat2016deep} - to identify how individual words are syntactically related in sentences. Syntactic connections are also enriched with synonym relationships - indicating which words can overlap in meaning according to WordNet 3.0~\citep{miller1998wordnet} - and emotional data - indicating how words were positively/negatively/emotionally perceived in validated psycholinguistic experiments~\citep{mohammad2013,mohammad2016sentiment}. The resulting multi-layer, feature-rich network structure contains insights about how text authors organised their knowledge and perceptions in texts. Through the lens of cognitive network science~\citep{siew2019cognitive,kenett2017semantic,stella2021cognitive}, discourse content, centrality and frames can all be measured via interpretable network metrics and, importantly, visualised. These aspects provide experimenters direct access into the structure of stances expressed in texts~\citep{stella2020forma} and also in the psychology of text authors~\citep{jackson2020text,teixeira2021revealing}. A key advantage of TFMNs is their ability to unveil semantic frames surrounding specific concepts in discourse as network neighbourhoods of concept associations. This is a crucial methodological feature for investigating specific aspects of phenomena as complex as vaccination campaigns.

\section{Methods}
\label{section_methods}

\subsection*{Data collection}
\label{subsection:data_collection}

We collected 5745 news articles about vaccines that circulated in Italy in a time window that spans from October 2020 to May 2021, along with their number of shares on social networks as Twitter and Facebook. We divided the dataset in articles coming from \textit{mainstream} media sources and \textit{alternative} sources. Mainstream sources are three among the most visited Italian newspapers, while alternative sources include a list of blogs, underground information and pseudo-newspapers already blacklisted by fact checkers of Bufale.net as persistent spreaders of mis- and disinformation. It must be noted, however, that none of the news in the dataset was verified by fact checkers nor by us. Thus, the two source categories are not indicative of true versus false contents, but rather of newspapers that have previously earned trust from the public opinion as authoritative and credible, versus unreliable and possibly partisan forms of news outlets. The full list of news sources is reported in Table~\ref{table:sources} together with their categorisation. The news selection was initially operated through the Twitter APIs, by retrieving tweets that contained a word related to vaccines and a url from the above list.  We considered to be related to vaccines keywords such as the Italian word ``vaccino'' itself, plus all the names of vaccines available worldwide by the time we collected data, i.e., ``pfizer'', ``astrazeneca'', ``vaxzevria'', ``moderna'', ``johnson'', ``sputnik'' and ``sinovac''.  We then scraped the articles, downloading the date, the title and the text content of the news. After discarding miscast articles, cancelled articles, articles protected by a paywall, and articles that were dated before the observation period (and merely re-tweeted in the observation period), we collected in total 3447 news from mainstream sources and 2298 from alternative sources. The number of shares on Twitter for a single article was inferred by the number of tweets we downloaded. Due to the limited amount of data retrieved and the current Twitter’s policies about APIs and rate limits, we can safely assume that our procedure downloaded most, if not all, the tweets that responded to the criteria introduced above. Last, we inferred for each url the number of shares on Facebook trough the tool for monitoring social media reactions Sharescore~\citep{sharescore}. Fig.~1 (b) shows a hint of the efficacy of the data retrieval procedure, displaying a good level of correlation between number Twitter and Facebook shares.

\begin{table}[]
\centering
\begin{tabular}{| c | c |} 
 \hline
 \textbf{Domain} & \textbf{Type}\\
 \hline
ilfattoquotidiano.it & Mainstream \\
repubblica.it & Mainstream \\ 
lastampa.it & Mainstream \\
\hline
imolaoggi.it & Alternative \\ 
voxnews.info & Alternative \\ 
renovatio21.com & Alternative \\ 
byoblu.com & Alternative \\ 
maurizioblondet.it & Alternative \\ 
scenarieconomici.it & Alternative \\
mag24.es & Alternative \\ 
irresponsabile.com & Alternative \\ 
disinformazione.it & Alternative \\ 
internapoli.it & Alternative \\ 
centrometeoitaliano.it & Alternative \\ 
essere-informati.it & Alternative \\
dionidream.com & Alternative \\ 
fonteverificata.it & Alternative \\ 
 \hline
\end{tabular}
\caption{List of the media outlets web domains analysed.}
\label{table:sources}

\end{table}

\subsection{Emotion detection and analysis of semantic frames}
\label{subsection:emotion_detection}

The main goal of this work is the analysis of the emotional fingerprint of news articles about vaccines and related concepts. Generally speaking, this task is performed by checking the texts against a lexicon of word-emotion associations (the NRC Lexicon~\citep{mohammad2013crowdsourcing}), but we included into our methodology two additional steps that increase the sensitivity and the significance of the analyses. 

First, we pre-processed all texts and extracted their \textit{Textual FormaMentis Networks} (TFMNs)~\citep{stella2020forma}. TFMNs, as we can see in Fig~1(a), provide a method for determining meaningful relationships between words, allowing for a fine-grain analysis about a single concept: by extracting the neighbourhood of a word from the TFMN of a (collection of) article(s), we were able to identify the words specifically associated to our target in the texts, filtering out words that merely co-occur in the same text but that have not a direct semantic or syntactic link with the target. 
Newspapers' texts can cover a wide spectrum of emotions, due to the length and the variety of subjects and facts within the same article. For instance, while talking positively about the impact of vaccines on daily casualties rates, an article could convey negative emotions about the deaths, or concerns about the future evolution of the outbreak. Overall, positive or negative framing of a concept may be diluted into the numerous traces of different emotions, thus being of primarily importance to explore semantic frameworks of words. 

To do so, we extracted the emotion distribution of the words belonging to the TFMN neighbourhood of a concept, by checking them against the Italian translation of the NRCLex lexicon. Last, we compared the emotions in the semantic frame to a null model, i.e., a random selection of words and their associated emotions. We computed the z-scores of the distribution of emotions that we found in TFMNs against the emotion distribution of 300 random samples of the lexicon itself. This methodology yields a numeric score for each emotion, which is an indication of how much that emotion is under-represented or over-represented. Fig.~\ref{fig:vaccin_daily}(D) and Fig.~\ref{fig:deathtromb}(C) have been generated using a slight modification of the visualisation library PyPlutchik~\citep{pyplutchik}, which allows for a quantitative representation of the Plutchik's wheel of emotions. Petals were sized after the z-score of how much an emotion has been detected in the TFMNs against a neutral baseline, and coloured only when they were at greater than 1.96 (or lower that -1.96), making it simple to identify emotions significantly over/under-expressed. A grey shadowed ring in such plots represents the space within 1.96 standard deviations from the average, where z-scores are not statistically odd. Similarly, in Fig.~\ref{fig:vaccin_daily}(C), we coloured only the lines representing those emotions that were below -1.96 or above 1.96 standard deviations from the average at least 50\% of the time, again emphasising odd patterns that were consistent over time.

\section{Results}

Results are organised along the following timeline. We first start by providing evidence that social engagement is not enough to identify differences between mainstream and alternative sources of information relative to COVID-19 vaccines. In other words, users tend to post about alternative and mainstream news at similar rates. However, the emotional content of these two sources of news differ drastically in the way they frame the idea of vaccines. Subsequently, we focus our attention on perceptions about specific vaccines like AstraZeneca. Lastly, we outline how news media reported negative concepts related to vaccines, like ``mort'' (stem for the Italian for death) and ``trombos'' (thrombosis).


\subsection{Prevalence of the discourse about vaccines on social networks and peaks of activity.} 
\label{subsection:prevalence}

Vaccines have been extensively debated both on social and on news media. Keeping track of the popularity of the articles reshared on social media can disclose insights about the ways mainstream and alternative news media described vaccination campaigns.
Figure 1(b) displays a correlation heatmap of the popularity of posts on the same set of news on Facebook and Twitter (see Methods), namely articles mentioning ``vaccin'' in their titles. It can be observed a positive correlation among the two measures of popularity (Pearson’s coefficient 0.38, p-value $\ll 0.001$). The correlation vanishes for extremely popular content, which is quite rare, and for URLs unpopular on Twitter, which can have diverse outcomes on Facebook. The densest mass in the plot is in the middle, where news are shared around 14 times on Twitter and 300 to 600 times on Facebook, confirming an order of magnitude of shift between the volume of the two social networks. Further details on social media users posting activity about vaccine-related news are provided in the Supplementary Information. Roughly the same positive correlation between Facebook and Twitter resharings persisted for both mainstream (Pearson’s coefficient 0.44, p-value $\ll 0.001$) and alternative (Pearson’s coefficient 0.41, p-value $\ll 0.001$) news outlets. This lack of differences indicates that both mainstream and alternative news were re-shared in similar ways across social media platforms. This also means that differences between mainstream and alternative news is to be searched not only within their online spread but rather in their semantic and emotional content, as investigated in the following subsection.

\subsection{Differences and similarities in the narrative about vaccine in mainstream and alternative news media articles}

Unable to distinguish between alternative and mainstream news, it has to be underlined that user activity on social media provides also little to no information on the narrative in which vaccines are framed. Hence, we rather focus our attention on quantifying the semantic/emotional content of news. To this end, we extract the semantic frameworks of selected concepts following a working pipeline explained in Figure~1(a) and in Sec.~\ref{section_methods} and, for starting, we explore the semantic framework of the term ``vaccin''.

Figure 1(c) reports that different kinds of media framed the concept of ``vaccin'' with wildly different emotions. In fact, mainstream Italian news media expressed significantly more trust and anticipation, as well as less disgust, than expected at random (see Methods). Such richly emotional framing of COVID-19 vaccines is completely absent in alternative sources. Whereas mainstream sources consistently framed ``vaccin'' with mostly positive/trustful jargon for more than 50\% of our sampling time window, alternative news framed the same concept as emotionless. Our results provide strong evidence that it is not social engagement but rather emotional profiling that strongly characterises sources coming from alternative and mainstream outlets.

However, Fig.~\ref{fig:vaccin_daily}(c) considers both titles and news bodies. Would these differences persist in case we focused on either of these elements? More importantly, would these differences be shared by specific types of vaccines like AstraZeneca?
 
In  Fig.~\ref{fig:vaccin_daily}(d) we focus on the AstraZeneca vaccine. Petals without colours in Fig.~\ref{fig:vaccin_daily}(d) represent emotions that are not significantly over- or under-expressed in the corpus, if compared with a neutral baseline (See. Section~\ref{section_methods}). We analyse the time-aggregated  emotions that are expressed on this noun over the whole time window, always referring to a neutral baseline. It is also interesting to differentiate not only between mainstream and alternative sources, but also among bodies and titles of the articles, since article headlines are meant to convey information and to catch the reader's eye in a very limited number of words. Using \textit{Plutchik's flower} to represent the eight primary emotions of the model, we find interesting differences. While both mainstream and alternative media adopt for their narratives the same emotional structure in the main bodies of their articles, different choices were made for the titles. Alternative sources expressed way more sadness compared to the baseline, while mainstream ones adopt a more positive attitude, expressing trust and anticipation.

\subsection{Perception of vaccine-related dangers: Before and after March 15, 2021}

We can further investigate the emotional approaches to vaccines of mainstream and alternative news media by studying the narratives before and after 15 March 2021. On this day the administration of the VaxZevria (then AstraZeneca) vaccine was temporarily suspended due to a small number of suspect side-effect thrombosis\footnote{https://www.aifa.gov.it/-/aifa-sospensione-precauzionale-del-vaccino-astrazeneca, Last Accessed: 7/12/2021}. An important media fuss was raised over this event, since the AstraZeneca vaccine was, at that time, one of the most used vaccine in Italy on several public worker categories\footnote{According to the National Vaccination Guidelines valid at that time, that can be found at the following URL (Italian only): shorturl.at/cBNT0, Last Accessed: 9/12/2021}. To do so, we conduct a two-fold analysis by analysing the semantic frameworks around the words ``pfizer'' and ``astrazenec'' through TFMNs, as well as around the words ``mort'' (death) and ``thrombos'' (thrombosis).

\begin{figure*}[ht!]
    \centering
    \includegraphics[width=1.00\textwidth]{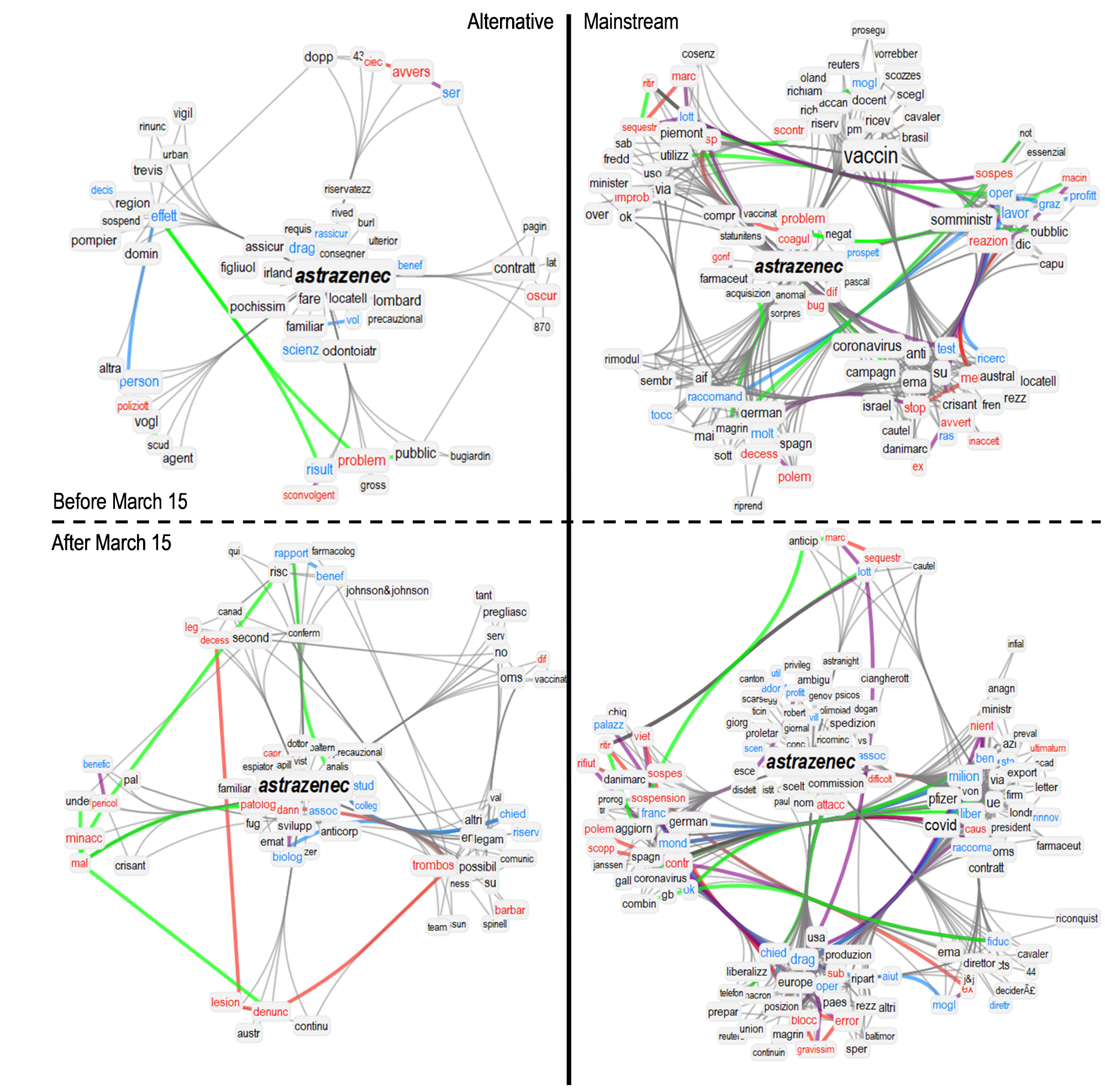}
    \caption{Semantic frames of tightly linked concepts around ``astrazenec'' in journal news titles from mainstream sources (right) and alternative sources (left), before (top) and after (bottom) the temporary suspension of 15 March 2021.}
    \label{fig:astratfmn}
\end{figure*}

\begin{figure*}[ht!]
    \centering
    \includegraphics[width=1.00\textwidth]{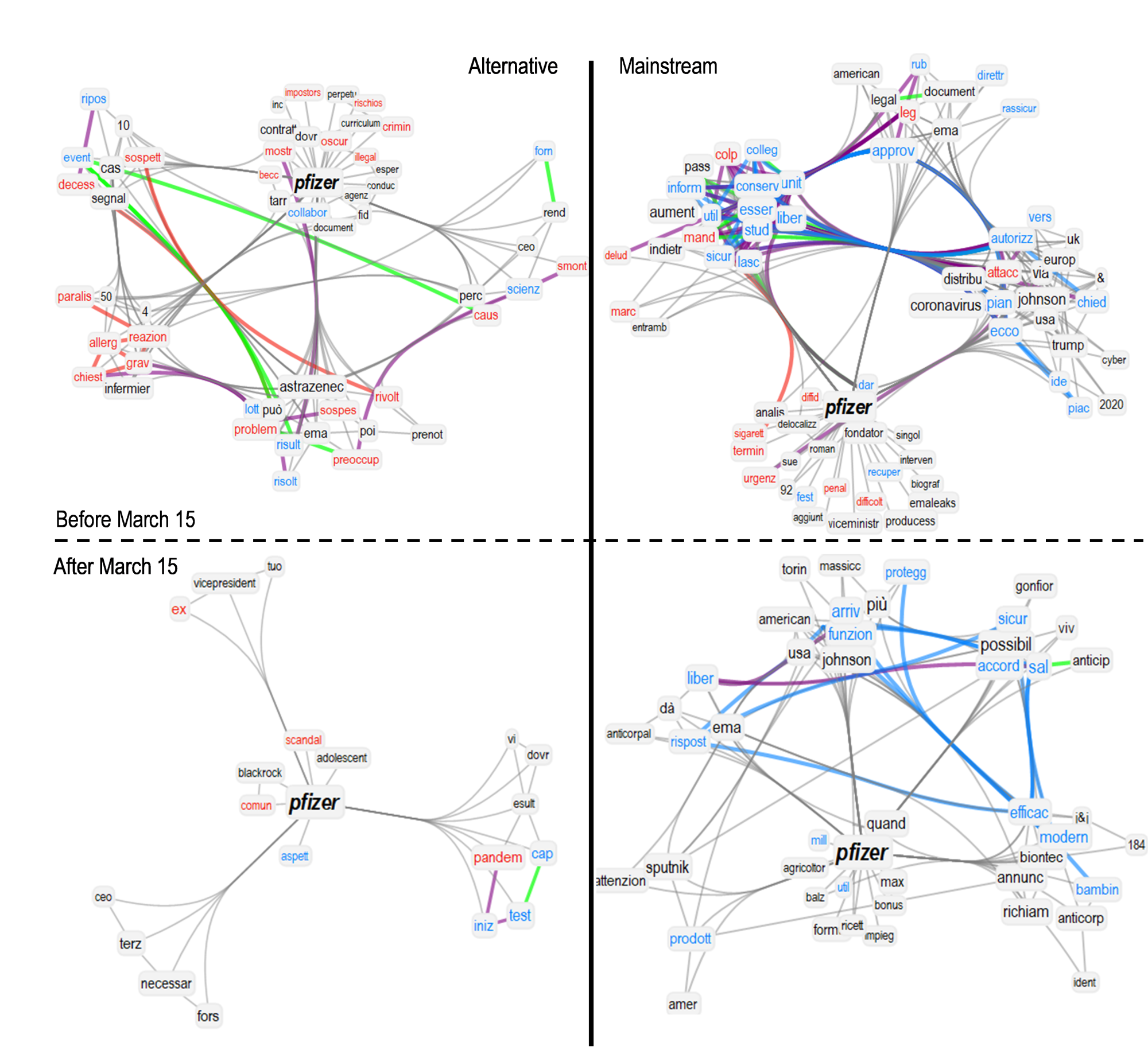}
    \caption{Semantic frames of tightly linked concepts around ``pfizer'' in journal news titles from mainstream sources (right) and alternative sources (left), before (top) and after (bottom) the temporary suspension of 15 March 2021.}
    \label{fig:pfitfmn}
\end{figure*}

Figures \ref{fig:astratfmn} and \ref{fig:pfitfmn} display the communities of tightly connected concepts surrounding, respectively, ``astrazenec'' and ``pfizer'' in their semantic frames as reconstructed from the TFMN. To highlight community structure we used the Louvain algorithm~\citep{blondel2008fast} as in previous works~\citep{stella2020forma}. Clustering concepts together in communities can better highlight more tightly connected and thus more semantically related concepts as debated in texts~\citep{stella2020forma}. These network visualisations report the semantic content associated with the above two entities in the corpus of news titles. Each network visualisation compared titles from mainstream sources (right) and alternative sources (left), before and after (top and bottom) the temporary suspension of VaxZevria on March 15th 2021.

Investigating the semantic content of these networks reveals interesting insights. Before the suspension, alternative journals framed ``astrazenec'' with way less negative associations than ``pfizer''. These journal venues concentrated negative associations like \textit{allergies}, \textit{reactions} and \textit{deaths} mostly when reporting about Pfizer's vaccine (cf. \ref{fig:pfitfmn} top left). These associations are absent in the titles surrounding ``astrazenec'' (cf. \ref{fig:astratfmn} top left). This indicates that alternative journal venues produced titles giving more semantic prominence to allergic/negative reactions to the vaccines mostly when mentioning Pfizer and more rarely when mentioning ``astrazenec''. This pattern is flipped when considering titles from mainstream journal venues, which feature more negative, reaction-related jargon when mentioning ``astrazenec'' (cf. \ref{fig:astratfmn} top right) rather than when talking about ``pfizer'' (cf. \ref{fig:pfitfmn} top right). On top on this disparity in reporting negative side effects of different brands of vaccines, titles from mainstream journals framed Pfizer with a positive cluster of concepts that is missing from AstraZeneca's semantic frame and relative to trust spawning from Pfizer's approval from experts and institutions.

After the temporary suspension of AstraZeneca, the semantic frames of ``astrazenec'' and ``pfizer'' underwent some drastic changes in the TFMNs obtained from news titles. Pfizer underwent a drastic drop of network degree (-79\%) in the titles from alternative journal venues, indicating a reduced semantic richness of language surrounding ``pfizer'' in those titles and, consequently, a reduced semantic prominence of Pfizer's vaccine in such titles (cf. \ref{fig:pfitfmn} bottom left). Always within titles coming from alternative news media, the semantic community of ``astrazenec'' underwent a \textit{densification of negative associations}. This included links with \textit{thrombosis}, \textit{threat} and \textit{dangerous} that were not present before. Mainstream journals produced semantic communities framing ``astrazenec'' in similar ways before and after the temporary suspension of the vaccine. Noticeably, mainstream venues featured associations with clusters of concepts related to bureaucracy, underlining how the vaccine was under further scrutinise.

\begin{figure*}[ht!]
    \centering
    \includegraphics[width=0.9\textwidth]{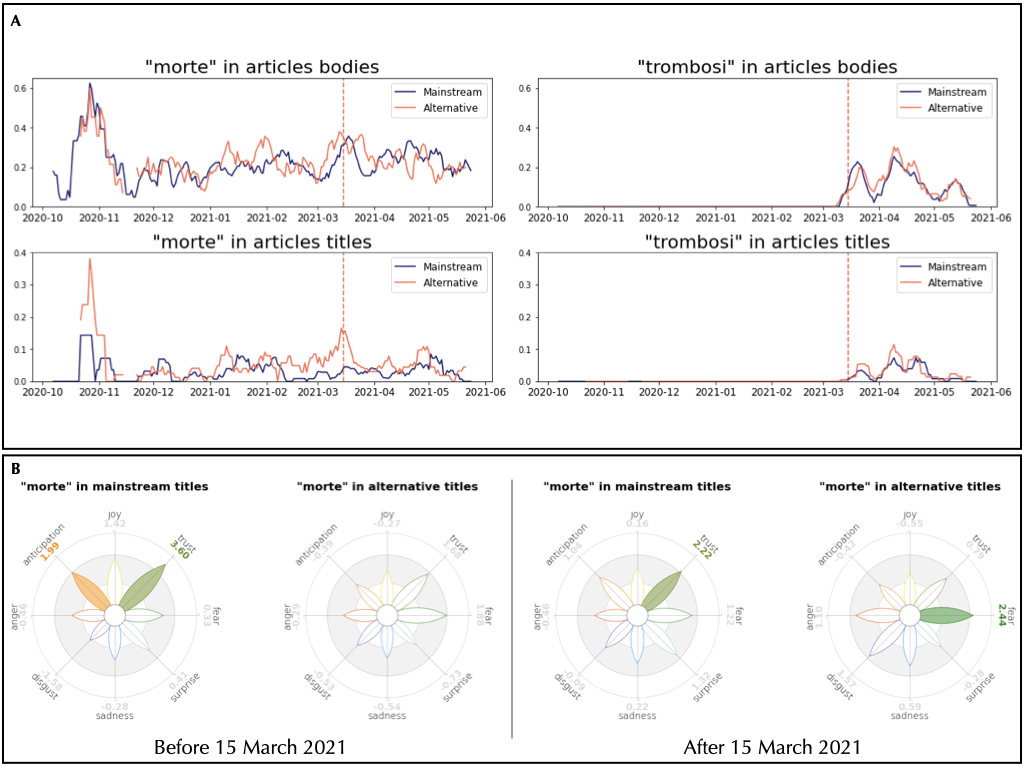}
    \caption{(A) Fraction of news that mention ``mort'' (death) and ``trombos'' (thrombosis) in their bodies and titles over the total of mainstream news (blue) and alternative news (red), smoothed by a moving average of the last 7 days. (B) Distribution of emotions in the semantic frame of ``mort'' (death) in the titles of mainstream and alternative articles.}
    \label{fig:deathtromb}
\end{figure*}

A further analysis driven by the events of March 15 can be conducted by tracking the usage of words associated to the negative events: \textit{death} and \textit{thrombosis}, the main alleged side effect that was put under the spotlight. An analysis of the prevalence of these two words, as shown in Fig.~\ref{fig:deathtromb}(a), suggests us that both mainstream and alternative media covered these events in a similar way.  The words ``mort'' (death) and ``trombos'' (thrombosis) have a long history due to its association with the COVID-19 outbreak, especially during the peak of casualties in Fall 2020, in both mainstream and alternative news outlets. The two plots show how many of the news produced by mainstream and alternative outlets everyday included the words ``mort'' (death) and ``trombos'' (thrombosis) respectively. Daily values were averaged with a moving average of 7 days, reducing the noise and increasing the readability of the figures. Indeed, they both show a peak of the said words around or after March 15, mentioning \textit{death} in almost 35\% of their articles and \textit{thrombosis} in the 25\% (mainstream) and 30\% (alternative) of their articles.

\section{Discussion}
News media shape the public opinion over many topics, including general health-related issues~\citep{gollust2013political,gollust2019television} as well as specific aspects of the fight against the pandemic, such as vaccine hesitancy~\citep{catalan2020vaccine,carrieri2019vaccine}. Needless to say, during the last two years COVID-19 has been one of the major talking points in all kind of news media. With this regard, many studies have been conducted on both the nature of COVID-related news and the effects such news have on the audience. While many have already pointed out the burst in COVID-19 online misinformation~\citep{yang2021covid,briand2021infodemics}, it has also been shown that users' online activity related to COVID-19 is strongly driven by media coverage~\citep{gozzi2020collective}. Therefore, general mass media play a fundamental role in steering the unfolding of the social consequences of the pandemic: being constantly exposed to COVID-related news might impact not only the pandemic itself, modifying the people's behaviour, but also the mental health of the audience~\citep{su2021mental}. This is a crucial aspect that should not be neglected: fear and anxiety can affect the general audience as well as those who are first in line in the response to the pandemic~\citep{coelho2020nature} and, particularly, using caution relatively to the COVID-19 media coverage is advisable~\citep{coelho2020nature}. This is especially true when considering that social networks, which as we saw are an important channel to share news articles, are a powerful tool for emotional contagion, that can also happen without direct interaction between people~\citep{kramer2014experimental}.

Indeed we found that, within the time window we considered, vaccines have been consistently portrayed with significantly more trust and anticipation in mainstream news, with no significantly emotional language displayed in alternative news. 
This general feeling towards the vaccines in alternative news was not the same reserved to the AstraZeneca vaccine for which, overall, carries significantly more sadness. This propensity for negative emotions in alternative news is reinforced as soon as we focus on the analysis around the date of 15 March 2020. This day has become a milestone in the Italian - and European - vaccination program, since the administration one of the most popular vaccines (AstraZeneca) underwent a series of disruptions due to a small number of serious side effects possibly related to the vaccine. Here, the usage of danger-related words such as \textit{death} or \textit{thrombosis} exploded in both mainstream and alternative media outlets, as did the emotional load, but it was in the latter where negative emotions prevailed, with a significant amount of fear dominating over the others. The analysis of the semantic neighbourhood of key concepts before and after the 15 March 2021 confirms that, after that date, there was a shift in the media's attention from the vaccine Pfizer to AstraZeneca, that was framed under an increasingly negative light, with new associations with negative words. 

Overall, we can appreciate significant differences between mainstream and alternative media sources. It is important to note though, that both kind of outlets insisted in presenting some topics (especially the AstraZeneca vaccine) in a strongly emotional way, with an almost always significantly higher emotional load with respect to a neutral language baseline. This is particularly true for the articles' titles: since they have to convey a short, effective message, they are - predictably - more loaded with emotional content than the articles' bodies, where the differences tend to taper. The media coverage of vaccines, throughout the whole pandemic, was emotionally intense; in a moment of crisis, such was the 15 March 2021, the media responded with a further diversification of the emotions, always higher than the neutral baseline. This fosters the findings presented at the beginning of this Section, where the authors highlight the strong interplay between media coverage of events and the mental health and response of the audience. Particular attention should be put both by the audience, for a well-reasoned consumption of media content, and by the media outlets for a careful choice of the narrative under which to describe sensitive matters, as are the events related to the vaccination campaign during a global pandemic. Given the proven influence of media coverage on vaccination campaigns~\citep{yoo2010effects,ma2006influenza,tchuenche2011impact}, conveying positive emotions, such as \textit{trust} and \textit{anticipation} whose combination, according to Plutchik's model~\citep{plutchik2001nature}, can be seen as an expression of \textit{hope}, could potentially have positive effects. Indeed, at the present date Italy shows an above-the-average adoption rate of vaccines against COVID-19, with 81\% coverage of the total population, performing better than United Kingdom (\%76), United States of America (\%74) and the average of the European Union (73\%)\footnote{According to https://ourworldindata.org/covid-vaccinations, last accessed: 12/01/2021}. On the other hand, an excess in optimism can also yield negative drawbacks~\citep{mccoll2022people}, proving once again how delicate is the balance in crisis communication and how important is the role of mass media in its management.

\subsection{Limitations and Future Work}

The present work has a number of limitations that should be taken into account while interpreting the results. The main limitation lies in the number of third-party resources used throughout all the study. Particularly, the URLs of the news media articles analysed have been collected by tracking down a selection of web domains and their diffusion on Twitter. These domains were flagged by independent annotators as either mainstream sources or misinformation spreaders, as per Table~\ref{table:sources}; to collect the URLs, we resorted to the Twitter APIs. This procedure brings two inherent limitations, related to both how up-to-date the independent lists are and how representative is the portion of tweets retrieved through the APIs. As for the former issue, we made sure that the references used to identify mainstream and alternative news media, commonly used by many other studies about Italian infodemics, were as current as possible at the moment of the data collection. 

Thanks to TFMNs we were able to avoid a further limitation that we could have encountered by quantifying emotions with a mere count of emotional words, referring to an external lexicon (such as the NRC Lexicon~\citep{mohammad2013crowdsourcing} that we used). Indeed, by applying this method - either to articles or to smaller excerpts, like titles or sentences - we would have likely lost sight of the context in which the words are used. TFMNs are instead a way to reconstruct such context, taking into account syntactic and semantic relationships between words. Such a results could be achieved by means of other NLP techniques, such as attention-based deep neural networks, more and more often used for emotions detection in texts~\citep{acheampong2021transformer}. Comparing the results yielded by these different methods could be very interesting, and a possible future research direction; as of now, we preferred to adopt a simple, light and - most importantly - entirely explainable approach, also in light of the sensible topic under discussion during these troubled times.

From a psychological perspective, it is important to underline that the current method highlights only emotions as described in text and not as felt by individual readers~\citep{stella2021cognitive}. This difference means that our study cannot, on its own, determine the consequences of being exposed to specific emotional profiles, e.g. did the negative associations attributed to "Pfizer" elicit vaccine hesitancy? Future research should be devoted to measure specifically how exposure to specific semantic/emotional content influences ways of thinking and, subsequently, behaviour in individuals.

\section{Conclusions}

News diffused via social media can portray the same event in different ways. Focusing on over 5000 Italian articles, re-shared more than 5 million of times on Facebook and Twitter, we show that \textit{mainstream} and \textit{alternative} news outlets framed COVID-19 vaccines in different ways. Whereas mainstream media adopted an overall positive stance towards vaccines, alternative outlets adopted less optimistic and more neutral semantic/emotional frames. This phenomenon is even more pronounced when it comes to specific pharmaceutical brands, particularly Pfizer and AstraZeneca, with news titles about the latter being characterised by strong \textit{sadness} (in alternative news) as opposed to the presence of hope (in mainstream titles). Our results also highlighted how emotions evolved within the narrative about COVID-19 vaccines in the press: after a crisis like the temporary suspension of the AstraZeneca vaccine in March 2021, words like \textit{death}
and \textit{thrombosis} came more frequently into play when debating vaccination campaigns. Supported by artificial intelligence, network science and psychologically validated emotional norms, the framework we introduced here provides a powerful tool for automatically unveiling emotional/semantic features of news content in large datasets. This tool opens the way to measuring differences in the way news media portray news as crucial and delicate as vaccination campaigns during a global pandemic.



 \bibliography{biblio}

\section*{Appendix: Supplementary Information}\label{sec:SI}

We first showcase some general, descriptive statistics about the prevalence and popularity of the discourse about vaccines on social media. As explained in Sec.~\ref{subsection:data_collection}, when collecting articles we distinguished between \textit{mainstream} and \textit{alternative} media outlets, having set the objective of finding whether there are significant differences in the narrative around vaccines between the two categories.

\begin{figure}[ht!]
    \centering
    \includegraphics[width=0.75\textwidth]{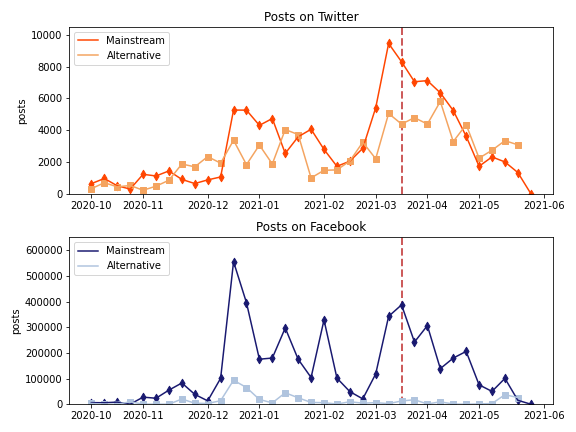}
     \caption{Posts on Twitter and Facebook in the dataset. A peak is registered in week 25 of observation, around the suspension of AstraZeneca on March 15.}
    \label{fig:posts_social}
\end{figure}

Figure~\ref{fig:posts_social} reports the amount of shares on Twitter (top) and Facebook (bottom), divided by mainstream and alternative media sources, and grouped by week. The activity on the two social networks is substantially different in numbers, with an average of 153 thousand shares by week on Facebook and only 5 thousands on Twitter, but it shows some resemblances. The number of daily posts slowly rises in late 2020, when the first news about the trial of Pfizer and AstraZeneca appeared; it bumps at the end of December 2020, when first Pfizer was approved for subministration\footnote{Source: https://www.aifa.gov.it/-/autorizzato-il-vaccino-biontech-pfizer, Last Accessed: 2/12/2021.}, and later at the end of January 2021, when AstraZeneca (now VaxZevria) was also approved\footnote{Source: https://www.aifa.gov.it/-/aifa-autorizzato-vaccino-astrazeneca, Last Accessed: 2/12/2021.} by AIFA, the Italian Agency for Drugs. The official press releases from AIFA are responsible for outstanding peaks of activity on Facebook (bottom), exactly around the approval of both vaccines. The number of posts diminishes in February, but it quickly rises to the highest level on March, around the suspension of AstraZeneca of March 15 (red dotted vertical line) due to a small number of suspect cases of thrombosis in Europe\footnote{https://www.aifa.gov.it/-/aifa-sospensione-precauzionale-del-vaccino-astrazeneca, Last Accessed: 7/12/2021}. While consistently higher than before, the volume of the discussion around both vaccines decreased steadily after March 15. While the general trends of engagement are similar on both social networks, there is a striking difference on what kind of sources drove the conversation about vaccines. The shares of news published by alternative sources can be considered negligible on Facebook, but on Twitter such news represent a massive quota of the total shares. To give a numeric comparison, total shares of alternative sources contents on Facebook were 8.36\% of the total, while on Twitter 43.61\%. Notice that Facebook counts might be underestimated because of sampling issues (see Methods) that originate from alternative sources being spread in smaller social pages that can go undetected by our external data gathering tools. On Twitter, where user activities are transparently monitored by the Twitter API (i.e., no third-party software) data exhibit no evident posting differences between alternative and mainstream news sources. The same high level of agreement is consistently present in mainstream news across Facebook and Twitter.





\end{document}